# ADC-Net: An Open-Source Deep Learning Network for Automated Dispersion Compensation in Optical Coherence Tomography

**Shaiban Ahmed[1], David Le[1], Taeyoon Son[1], Tobiloba Adejumo[1] and Xincheng Yao[1,2]***

[1]Department of Biomedical Engineering, University of Illinois Chicago, Chicago, Illinois, USA.

[2]Department of Ophthalmology and Visual Science, University of Illinois Chicago, Chicago, Illinois, USA.

**\*Corresponding Author:** xcy@uic.edu

**Abstract:** Chromatic dispersion is a common problem to degrade the system resolution in optical coherence tomography (OCT). This study is to develop a deep learning network for automated dispersion compensation (ADC-Net) in OCT. The ADC-Net is based on a redesigned UNet architecture which employs an encoder-decoder pipeline. The input section encompasses partially compensated OCT B-scans with individual retinal layers optimized. Corresponding output is a fully compensated OCT B-scans with all retinal layers optimized. Two numeric parameters, i.e., peak signal to noise ratio (PSNR) and structural similarity index metric computed at multiple scales (MS-SSIM), were used for objective assessment of the ADC-Net performance. Comparative analysis of training models, including single, three, five, seven and nine input channels were implemented. The five-input channels implementation was observed as the optimal mode for ADC-Net training to achieve robust dispersion compensation in OCT.

## 1    Introduction

Optical coherence tomography (OCT) is a non-invasive imaging modality that can provide three-dimensional information for clinical assessment (1, 2). By providing micrometer scale resolution to visualize retinal neurovasculature, it has been widely used for ophthalmic research and clinical management of eye conditions (3-5). Given the reciprocal relationship between the axial resolution and bandwidth of the light source, high resolution OCT requires a broadband light source (6, 7). However, the broadband light source induces chromatic dispersion, i.e., light wavelength dependence of the optical pathlength difference between the sample and the reference arms. The reference arm generally houses a reference mirror that has a uniform reflectance profile, but the sample arm usually contains dispersive media (such as a biological tissue) and when light travels through such medium, the different wavelength components correspond to different optical path lengths. This induces phase shifts among different wavelength signals in OCT detection, and thus degrades the axial resolution. Dispersion can also produce chirping noise and hence reduce the quality of the OCT image. Both hardware and numeric methods have been developed for dispersion compensation to enhance OCT image quality.

The hardware-based method involves additional dispersive media, such as water, BK7 glass (8), fused silica (9), etc. in the reference arm to balance the dispersion in the sample arm. Physical dispersion compensation can also be achieved by using grating-based group and phase delay detectors (10). Faster speed and a higher range of tunability were achieved by using acousto-optic modulator (11) and tunable optical fiber stretchers (12). However, hardware compensation leads to a bulky optical system due to the installation of additional components. Moreover, the hardware compensation is typically effective only when the sample subject is stable with a fixed dispersion.

Numerical dispersion compensation has been established as a useful alternative to hardware-based techniques. Numeric dispersion correction is based on the digital processing of OCT spectrogram, providing the flexibility of adjustable correction values of the dispersion induced phase error. Fercher et. al. (13) proposed a numeric compensation technique where a depth-dependent kernel was correlated with the spectrogram to compensate dispersion. However, this method relies on the availability of information related to the dispersive characteristics of the sample which can vary for biological tissues. Fractional Fourier transform for dispersion compensation was introduced by Lippok et al. (14) where the performance is dependent on the accuracy of the value of an order parameter and acquiring this value for biological samples can be challenging. A spectral domain phase correction method, where the spectral density function is multiplied with a phase term, was demonstrated by Cense et. al. (15). However, to determine the precise phase term, isolated reflections from a reference interface with a uniform reflectance profile, which might not be available in clinical setup, are required. A numeric algorithm based on the optimization of a sharpness function, which was termed to be divided by the number of intensity points above a specified threshold, was presented by Wojtkowski et. al. (16) where a pair of dispersion compensation coefficients were derived to compensate dispersion in the entire B-scan. But due to the depth varying changes in biological tissues, the dispersion effect can vary at different depths and hence a single pair of dispersion compensation coefficients may not be able to compensate dispersion for all depths effectively in one B-scan.

Entropy information of the signal acquired in the spatial domain was utilized as the sharpness metric by Hofer et. al. (17) to compensate dispersion. However, the numeric techniques which are based on sharpness metrics are susceptible to the prevalent speckle noise in OCT B-scans and can lead to overestimation or underestimation when the system lacks high sensitivity. A depth-reliant method was proposed by Pan et. al. (18) where an analytical formula was developed to estimate the second-order dispersion compensation coefficients in different depths based on a linear fitting approach. But this method relies on the accurate estimation of second-order coefficients at specific depths and the analytical formula can differ for different biological subjects. Besides, the lower degree of freedom available in a linear fitting method can lead to inaccurate estimation of the coefficients at different depths. Spectroscopic analysis of A-scan's spectrogram was conducted to estimate and correct dispersion by Ni et. al. (19) where information entropy estimated from a centroid image was used as a sharpness metric. However, this technique leads to lower resolution and requires a region of analysis without transversely oriented and regularly arranged nanocylinder. In general, these classical numerical methods can be computationally extensive when it comes to the widescale application as they are usually designed based on specific conditions and may require additional optimization for the generalized application. Hence these methods may lead to computational complexity in real-time application.

Deep learning has garnered popularity in medical image processing (20-24), with demonstrated feasibility for real-time application due to its capability to handle large datasets, computational efficiency, high accuracy, and flexibility for widescale application. Deep learning-based algorithms have been used in image denoising (25, 26), segmentation (27-31), classification (32-34), etc.. In this study, we propose a deep learning network for automated dispersion compensation (ADC-Net) that is based on a modified UNet architecture. Input to ADC-Net comprises OCT B-scans which are compensated by different second-order dispersion coefficients and hence are partially compensated for certain retinal layers only. The output is a fully compensated OCT B-scan image optimized for all retinal layers. We quantitatively analyzed the proposed model using two parameters namely MS-SSIM and PSNR. Comparative analysis of training models, including single, three, five, seven and nine input channels were implemented. The source code along with necessary instructions on how to implement it have been provided here: https://github.com/dleninja/adcnet

## 2 Material and Methods

This study has been conducted in compliance with the ethical regulations reported in the Declaration of Helsinki and has been authorized by the institutional review board of the University of Illinois at Chicago.

### 2.1 Data Acquisition

Five human subjects (mean age: $30 \pm 4.18$ years; mean refractive error: $-2.28 \pm 1.53$D) were recruited to acquire the OCT images for training and testing the proposed ADC-Net. These subjects had no history of ocular medical conditions. All human subjects affirmed their willful consent before participating in the experimental study. For OCT imaging, the illumination power on the cornea was 600 µW which is within the limit set by the American National Standards Institute. The light source used for this experiment was a near infrared (NIR) superluminescent diode (D-840-HPI, Superlum, Cork, Ireland). A pupil camera and a dim red light were used for localizing the retina and as a fixation target, respectively. The purpose of the fixation target is to minimize voluntary eye movements. Axial and lateral pixel resolutions were achieved as 1.5 µm and 5.0 µm respectively. The OCT spectrometer that was used for data recording consisted of a line-scan CCD camera. The total number of pixels in the CCD camera was 2048 pixels and the line rate was 70,000 Hz. The recording speed for OCT imaging was 100 B-scans per second with a frame resolution of 300 A-lines per B-Scan. A total of 9 OCT volumes were captured. Each OCT volume consists of 1200 B-scans. Seven of these OCT volumes (8400 B-scans) were used for training the model, and another two (2400 B-scans) were used as testing set.

### 2.2 Dispersion Compensation

The signal acquisition in OCT involves recording the spectrogram obtained by interfering back-reflected light from different interfaces of the sample with the back-reflected light from the reference mirror. The fringe pattern generated by this interference signal is detected by the spectrometer and corresponding OCT signal can be represented by the following equation (18):

$$S_{int}(k) = 2Re\left\{\sum\sqrt{I_n(k)I_r(k)}\exp\{i[k.\Delta z_n + \phi(k.\Delta z_n)]\}\right\} \qquad (1)$$

In this equation, $I_n(k)$ and $I_r(k)$ represent the back-reflected light intensity obtained from the $n$-th sample layer and reference arm respectively. The wavenumber is denoted by $k$ while the difference in optical path length is denoted by $\Delta z_n$. The phase term $\phi(k.\Delta z_n)$ denotes the phase difference between the sample interfaces and the reference mirror and this includes the higher order dispersive terms. We can express the phase term as:

$$\phi(k, \Delta z_n) = \beta_n(k) \cdot \Delta z_n$$

$$= \left[n_n(k_o) \cdot k_o + n_{g,n}(k_o) \cdot (k - k_o) + \beta''_n(k_o) \cdot \frac{(k - k_o)^2}{2!} + \beta'''_n(k_o) \cdot \frac{(k - k_o)^3}{3!} + \cdots\right] \cdot \Delta z_n$$

$$= n_n(k_o) \cdot k_o \cdot \Delta z_n + n_{g,n}(k_o) \cdot \Delta z_n \cdot (k - k_o) + a_2 \cdot (k - k_o)^2 + a_3 \cdot (k - k_o)^3 + \cdots \qquad (2)$$

Here, $\beta_n$ represents dispersion coefficient, while $a_2$ and $a_3$ are second and third-order dispersion compensation coefficients. While $n_n$ is the sample's $n$-th layer's refractive index, $n_{n,g}$ is the effective group refractive index. Numeric dispersion compensation can be done by modifying the phase term through the addition of a phase correction term which eliminates the dispersive phase. The following equation shows second and third-order dispersion compensation phase correction:

$$\overline{\phi}(k) = -a_2(k - k_o)^2 - a_3(k - k_o)^3 \tag{3}$$

Here, $a_2$ and $a_3$ can be adjusted to compensate second-order group velocity and third-order phase dispersion. However, since dispersion in biological tissue is different at different depths, dispersion compensation using a single pair of second and third-order coefficients might not be sufficient for all depths. Numerically estimating and applying different compensation coefficients for different depths can be computationally extensive for widescale application. In following section 2.3, we present ADC-Net, a deep learning-based compensation algorithm for automated implementation of full depth compensation. Input to ADC-Net can be of single or multiple channels of partially compensated B-scans. These partially compensated B-scans can be obtained by using the phase correction method in equation (3). For simplicity, we have compensated the B-scans using second-order compensation only. All depth dispersion compensated ground truth data were also crafted from an array of partially compensated B-scans and the detailed procedure is elaborated in section 2.4.

## 2.3 Model Architecture

The ADC-Net is a fully convolutional network (FCN) based on a modified UNet algorithm, which consists of an encoder-decoder architecture (Figure. 1). The input to the ADC-Net can be of a single channel or a multichannel system. Each input is an OCT B-scan image which was compensated by different second-order dispersion compensation coefficients and hence the B-scans in each channel are optimally compensated at different layers or depths. The output is dispersion compensated OCT B-scans where all layers in different depths are compensated effectively.

The encoder segment is a combination of convolutional, max pooling, dense, and transitional blocks. The primary function of the decoder segment is to deduce useful features from the image. To ensure precise feature localization and mapping for generating output images, bridging between the encoder and the decoder is established. The convolution blocks, which perform summing operations, constitute the dense blocks. The skip connections, which alleviate the vanishing gradient problem, are used to link each subsequent block to previous blocks. A transition block is connected to each dense block, which is to reduce the dimension of output feature map.

The decoder segment consists of up-sampling operations along with the decoder blocks. Using the decoder block, the outputs obtained from the convolution operation of the fitting transition blocks and the up-sampling operations are concatenated. Image features can then be localized precisely by convolving the generated feature maps.

In the ADC-Net, two types of functions, namely batch normalization function and ReLU activation function trail all the convolution operations. On the other hand, a SoftMax activation function follows the terminal convolutional layer.

Moreover, transfer learning is employed to avoid overfitting errors by utilizing the ImageNet dataset. The ImageNet dataset is a visual database that consists of millions of everyday images. These images differ from the OCT B-scans but facilitate in training the CNN model to learn about simple features such as edges, color codes, geometric shapes, etc. in the primary layers and complex features in the deeper layers by utilizing CNN's bottom-up hierarchical learning structure. Transfer learning can then facilitate the network to relay these simple features to learn complex features which are related to the OCT B-scans. A fully connected layer that consists of 1000 neurons along with a SoftMax activation function exists in the pre-trained encoder network model. When the pre-training was concluded (after achieving about 75% classification accuracy on the ImageNet validation dataset), the fully connected layer was removed. The decoder network was then fed with transitional outputs from the encoder network. Adam optimizer which had a learning rate of 0.0001 was used to train the FCN model along with a dice loss function.

The hardware environment used for implementing the proposed ADC-Net had Windows 10 operating system equipped with NVIDIA Quadro RTX Graphics Processing Unit (GPU). The software model was written on Python (v3.7.1) utilizing Keras (v2.2.4) with Tensorflow (v1.31.1) backend.

## 2.4 Data Pre-Processing

Figure. 2 briefly illustrates the ground truth preparation method for a single B-scan. Raw OCT data were acquired using the SD-OCT system described in section 2.1. An array of B-scans ranging from $I_1$ to $I_N$ (In this study, N = 5) were reconstructed from the same raw data frame using the usual procedure that involves background subtraction, k-space linearization, dispersion compensation, FFT, etc. However, each of the B-scans were compensated with different second-order dispersion compensation coefficients ranging from $C_1$ to $C_N$ (N = 5). Technical rationale of numeric compensation has been explained in Section 2.2. Since the tissue structure in a biological subject differ at different depths, the dispersion effect also varies accordingly and thus a single second-order coefficient can effectively compensate dispersion errors at a specific layer only. Thus, required values of second-order dispersion compensation co-efficient, ranging from $C_1$ to $C_N$, were selected empirically so that dispersion at all depths were compensated optimally. In $I_1$, the region demarcated by the red box along the inner retina had been dispersion compensated and optimized using $C_1$. However, as we move further away from the inner layers the dispersion effects appear to be more prominent due to ineffective compensation. Image $I_N$ on the contrary has the region demarcated by the yellow box at the outer retina optimized and well compensated. To prepare the ground truth B-scan, the red and yellow demarcated region from $I_1$ and $I_N$ were extracted and stitched in proper sequence to obtain dispersion compensated layers at the inner and outer retina. Similarly, the remaining layers acquired from $I_2$ to $I_{N-1}$ which were compensated by $C_2$ to $C_{N-1}$ respectively. Optimally compensated layers were extracted and stitched sequentially to obtain the all-depth compensated ground truth B-scan. To prepare the training and test data for the single, three, five, seven, and nine input channel models, 1, 3, 5, 7, and 9 arrays of B-scans were re-constructed respectively from each raw volume while each array were compensated with different second-order dispersion compensation coefficients. These coefficients were selected in equal intervals between $C_1$ to $C_N$. To acquire the OCT data and for digital image processing, LabView (National Instruments) and MATLAB 2021 software environments were used, respectively.

## 2.5 Quantitative Parameters

Two parameters, namely peak signal to noise ratio (PSNR) and structural similarity index metric at multiple scales (MS-SSIM) were used for quantitative analysis and objective assessment of our proposed method. The two parameters are defined as follows:

**Peak Signal to Nose Ratio:** PSNR can be defined as the ratio of maximum signal strength to the corrupting background noise which was computed using the following equation (35):

$$PSNR = 10log_{10}\left(\frac{s^2}{MSE(f,g)}\right) \qquad (4)$$

Here, $s$ is the maximum pixel intensity value in the reconstructed image. Mean squared error (MSE) between the reference image $f$ (ground truth) and reconstructed image $g$ (output) can be defined by the following equation:

$$MSE(f, g) = \frac{1}{MN} \sum_{i=1}^{M} \sum_{j=1}^{N} (f_{ij} - g_{ij})^2 \qquad (5)$$

Here, $M$, $N$ denotes the number of rows and columns while $(f_{ij} - g_{ij})$ denote the pixel-wise error difference between $f$ and $g$.

**Structural Similarity Index Metric at Multiple Scale:** MS-SSIM was computed to quantify the structural similarities between the ground truth and the corresponding output images obtained from the ADC-Net when implemented with different input channels models. MS-SSIM utilizes three visual perception parameters namely the luminance, contrast and structural parameters when calculated at multiple scales and thus it incorporates detailed image information at different resolutions and visual perceptions which make it a robust and accurate quality metric.

If $x$ and $y$ denote two image patches which are to be compared, the luminance parameter is defined by (36):

$$l(x, y) = \frac{2\mu_x \mu_y + C_1}{\mu^2_x + \mu^2_y + C_1} \qquad (6)$$

The contrast parameter is defined by:

$$c(x, y) = \frac{2\sigma_x \sigma_y + C_2}{\sigma^2_x + \sigma^2_y + C_2} \qquad (7)$$

The structural parameter is defined by:

$$s(x, y) = \frac{\sigma_x \sigma_y + C_3}{\sigma_x \sigma_y + C_3} \qquad (8)$$

Here, $\mu_x$ and $\mu_y$ represent the mean while $\sigma_x$ and $\sigma_y$ represent the standard deviation of $x$ and $y$ respectively. The constants $C_1$, $C_2$, and $C_3$ can be obtained by:

$$C_1 = (K_1 L)^2, \ C_2 = (K_2 L)^2 \quad and \quad C_3 = C_2/2 \qquad (9)$$

Where, L is the pixel dynamic image range and $K_1$ *(0.01)* and $K_2$ *(0.03)* are two scalar constants. Thus, SSIM can be defined as:

$$SSIM(x, y) = [l(x, y)]^\alpha \cdot [c(x, y)]^\beta \cdot [s(x, y)]^\gamma$$

$$or, SSIM(x, y) = \frac{(2\mu_x \mu_y + C_1)(2\sigma_x \sigma_y + C_2)}{(\mu^2_x + \mu^2_y + C_1)(\sigma^2_x + \sigma^2_y + C_2)} \ [\alpha = \beta = \gamma = 1] \qquad (10)$$

In order to obtain multi scale SSIM, an iterative approach is employed where the reference and the output images are scaled *M-1* times and down sampled by a factor of 2 after each iteration. The contrast and structural parameter are calculated at each scale while the luminance parameter is computed only at the *M*-th scale. The final quantitative parameter is obtained by the combining the values obtained at all the scales using the following relation:

$$MSSSIM(x, y) = [l_M(x. y)]^{\alpha M} \cdot \prod_{j=1}^{M} [c_j(x, y)]^{\beta j} [s_j(x, y)]^{\gamma j} \qquad (11)$$

## 3    Results

Figure 3A shows representative OCT B-scan images obtained from different input channels models along with a raw uncompensated OCT B-scan. For better visualization, 6 neighboring B-scans at the macula region of a human retina were averaged and the images are displayed on a logarithmic scale. Figure 3A1 represents the uncompensated image and due to the dispersion effect, the B-scan suffers from low axial resolution and the different retinal layers appear to be blurry and overlapping as the detailed structural information is lost. Figure 3A2-A6 illustrate the representative OCT B-scans obtained from single to nine input channels models respectively. In Figure 3A2, which was obtained from the single input channel model, even though the inner retinal layer appears to be well compensated, the central and outer bands suffer from blurring and the dispersion effect is not optimally compensated. When the input was increased to 3, 5, 7, and 9 B-scans, the quality of the output image was enhanced. As shown in Figure 3A3-A6, both the inner and outer layers were better compensated compared to single input channel model and showed sharp microstructural information. As described in section 2.1, input models with 3, 5, 7, and 9 input channels had input B-scans with both inner and outer retinal layers compensated. However, 5,7, and 9 input channels performed slightly better compensation compared to the model with 3 input channels and this improvement can be better visualized in Figure 3B which were generated for a detailed illustration of the differences in performance of the different models.  To generate these images, pixel-wise, intensity difference between the corresponding ground truth and Figure 3A1-A6 were computed, and the resultant intensity differential images were displayed in a jet colourmap. Here, the bright regions indicate higher differences in intensity while the darker blue region indicates lower to no difference. These images serve two purposes. First, the pixel wise extent of dissimilarity between each of the images and the ground truth can be observed. The lesser difference with the ground truth means higher similarity and thus indicates better performance. Second, the detailed differences in performance between the different input models and the uncompensated image can be better visualized using the difference between these images and the ground truth as a qualitative metric. Figure 3B1 shows that the uncompensated image depicts more difference from the compensated ground truth due to lack of dispersion compensation. While Figure 3B2 shows slightly better performance but more bright regions at the outer retina depict that the single input model could not compensate dispersion properly at the lower depths. Figure 3B3 illustrates that the model with three input channels performed better than the single input channel model. However, Figures 3B4-B6 show almost similar performance and the least amount of difference with the ground truth and thus the higher performance than the other two models.

The two quantitative parameters described in section 2.5 were calculated from the resultant images obtained from different input channels models along with the corresponding uncompensated images to perform a quantitative assessment of the proposed ADC-Net. Before computing the quantitative parameters, the four repetitive B-scans at the same locations were averaged. MS-SSIM and PSNR were calculated, and the result is graphically represented in Figure 4. Mean values with standard deviation were used for representative purpose. In Figure 4, UC represents uncompensated images while M1, M3, M5, M7, and M9 represent the output obtained from single, three, five, seven, and nine input channels models respectively. In Figure 4A, we can observe that the lowest mean MS-SSIM score was obtained for UC ($0.85 \pm 0.025$) which depicts the least similarity with the ground truth image. Due to the dispersion effect the image quality degrades significantly without compensation. The MS-SSIM score obtained from the single input channel model (M1) is $0.94 \pm 0.021$ which shows an improved performance in terms of dispersion compensation compared to the raw uncompensated image. The similarity score for three (M3) and five (M5) input channels models show a gradual improvement in performance with MS-SSIM values of $0.95 \pm 0.018$ and $0.97 \pm 0.016$ respectively. However, the graph flattens after M5 as seven (M7) and nine (M9) input channels models show a similarity score of $0.97 \pm 0.014$ and $0.97 \pm 0.014$ which are within the 1 standard deviation range of the five-input channels model. We can observe a similar trend in Figure 4B which depicts the mean PSNR. Highest PSNR of 29.95 ± 2.52 dB was calculated for the five input channels model (M5) while seven and nine input channels model had a mean PSNR of 29.91 ± 2.134 dB and 29.64 ± 2.259 dB respectively. Output from the three input channels model recorded a slightly lower PSNR value of 27.49 ± 1.606 dB and the downward slope continued for the single input channel model (M1) with a mean value of 25.86 ± 1.677 dB and the least PSNR of 20.99 ± 0.021 dB was observed for the uncompensated B-scans.

Figure 5 illustrates comparative reflectance intensity profile analysis of the outer retina. The yellow vertical line in Figure 5A shows the retinal region for OCT intensity profile analysis (Figure 5B). Figure 5B illustrates axial intensity profiles at the parafovea and encompasses the outer retinal bands. Six neighboring B-scans were averaged and 5 adjacent A-lines at the region of interest were averaged from the averaged B-scan before generating the intensity profiles. Figure 5A is displayed in logarithmic scale for enhanced visualization, but the intensity profile analysis in Figure 5B is shown in linear scale. The intensity profiles are depicted in Figure 5 where GT and UC represent the intensity profiles obtained from the ground truth and the uncompensated images, respectively. Whereas M1, M3, M5, M7, and M9 stand for the intensity profiles obtained from single, three, five, seven, and nine input channels models respectively.

The ELM band profile is known to reflect the point spread function (PSF), i.e., the axial resolution. From GT we can observe a sharper and narrower PSF at the ELM layer when compared to UC and M1, where the PSF is flat and wider. The ELM band profile becomes slightly better for M3, but M5, M7, and M9 depict thinner and analogous band profile to GT. Blurred RPE band profiles can also be observed for UC and M1 at the RPE which overlaps with the Bruch's Membrane (BM) region (37). On the contrary, GT, M3, M5, M7, and M9 show sharper peaks which can be distinguished separately. This means that the RPE and Bruch's membrane can be observed separately from the reconstructed images. The IS/OS and OPR bands in UC have distinguishable peaks but still depict thicker profiles, compared to GT. M1 and M3 show slightly thinner IS/OS and OPR bands, compared to UC. On the other hand, GT, M5, M7, and M9 depict sharper and thinner OCT band profiles.

Dispersion can also shift the location of the interfaces in a multilayered sample which can affect the depth measurement of different layers (38). We comparatively evaluate the peak locations of the ELM, IS/OS, OPR, and RPE in the intensity profiles demonstrated in Figure 5 to assess the performance of our proposed method in terms of depth measurement where the GT was taken as the reference of assessment. The GT's peak location at the ELM and IS/OS layer aligns with M3, M5, M7, and M9. However, the peak location for M3 shifts at the OPR and RPE when compared to the peak location of the GT where M5, M7, and M9 demonstrate aligned peaks with the GT. For M1 and UC, the peaks observed at ELM and RPE are flat and overlapping while the peaks are shifted at the IS/OS and OPR layer.

## 4    Discussion

Dispersion compensation is necessary to obtain high axial resolution and retain detailed structural information in OCT. Traditional numeric dispersion compensation approaches can be computationally expensive. Numerically devised methods also require optimization based on specific contexts, and thus may lack flexibility for generalized application. The demonstrated ADC-Net can be automated for real-time implementation due to its higher computational flexibility and simplicity. Once trained with the optimum number of input channels and well-crafted ground truth data, ADC-Net can automatically compensate dispersion effectively and generate OCT B-scans with high quality. We made the proposed ADC-Net available through an open-source platform (https://github.com/dleninja/adcnet) for easy accessibility to a robust and automated dispersion compensation algorithm.

The performance of ADC-Net peaked when employed using five, seven, and nine input channels models. While a single input channel model performed better than a raw uncompensated image in terms of image resolution, the output images still depicted blurring effects. The output obtained from the three input channels model was better than the single input channel model but slightly worse than the five, seven, and nine input channels.

Since the proposed FCN is built on a modified UNet structure and follows an encoder-decoder pipeline, the model trains itself by acquiring features from the input and the ground truth data to reconstruct dispersion compensated images. For a single input channel model, the input B-scans were compensated by a single second-order dispersion compensation co-efficient which can optimize dispersion in a specific retinal layer only. For our experiment, the second-order dispersion compensation parameter was selected close to $C_1$ and thus the input B-scans had optimum compensation along with the inner retinal layer only. Consequently, the output B-scans had the inner retinal layer optimized only. The input model with three input channels had three arrays of B-scans and each array was compensated by different second-order dispersion compensation coefficients which were selected equally spaced between $C_1$ to $C_N$. The three input channels thus provided more information related to more layers being compensated and thus the model performed better compared to the single input channel model. Similarly, for five, seven, and nine input channels models, the input channels had more B-scans with more layers being compensated which in terms provided more features to the model to train itself better. Hence the performance was better compared to single and three input channels models. However, quantitative analysis revealed that five, seven, and nine input channels models depict similar performance, and thus optimum all-depth dispersion can be obtained using five input channels.

The dispersion effect broadens OCT band profiles and thus degrade the axial resolution. In a well-compensated OCT image, such as the ground truth image, this blur effect would be minimized, corresponding to thinner and sharper band profiles. On the other hand, as illustrated in the intensity profile obtained from the uncompensated image, the band profiles would be thicker due to dispersion effect which in terms would affect the image resolution. This would impact the thickness measurement of the retinal bands as they would appear to be thicker than the actual value. From our proposed ADC-Net we obtained sharp and thin OCT band profiles from input models with five, seven, and nine channels which were analogous to the intensity profile obtained from the ground truth image. The peaks for the outer retinal layers were also aligned which shows the promise for accurate depth measurement. Hence the proposed ADC-Net demonstrates its capability to generate B-scans with high resolution that can retain intricate structural information. Implementation of this automated process can be beneficial in clinical assessment and ophthalmic research by providing accurate retinal thickness and depth measurement in healthy and diagnosed patients. Artificial intelligence may reduce the technical complexities and streamlining tasks in a clinical setting.

The major challenge of our proposed ADC-Net is the availability of finetuned ground truth data which requires all depth compensation. However, once the required values of the coefficients for different depths and ranges are obtained for one volume, it can be applied to other volumes directly. Thus, for one OCT system, calibrating the system once would be enough. Once our proposed ADC-Net is trained with all depth compensated ground truth, it can automatically generate fully compensated B-scans with retinal layers optimized. Therefore, this ADC-Net process can be effectively implemented for real-time application.

## 5    Conclusion

A deep learning network ADC-Net has been validated for automated dispersion compensation in OCT. The ADC-Net is based on a redesigned UNet architecture which employs an encoder-decoder pipeline. With input of partially compensated OCT B-scans with individual retinal layers optimized, the ADC-Net can output a fully compensated OCT B-scans with all retinal layers optimized. The partially compensated OCT B-scans can be produced automatically, after a system calibration to estimate the dispersion range. Comparative analysis of training models, including single, three, five, seven and nine input channels were implemented. The five-input channels implementation was observed to be sufficient for ADC-Net training to achieve robust dispersion compensation in OCT.

**Conflict of Interest:** The authors declare that the research was conducted in the absence of any commercial or financial relationships that could be construed as a potential conflict of interest.

**Author Contributions:** SA contributed to data processing, analysis, model implementation and manuscript preparation. DL contributed to network design, model implementation and manuscript preparation. TS contributed to experimental design, data acquisition, and manuscript preparation. TA contributed to data processing. XY supervised the project and contributed to manuscript preparation.

**Funding:** This research was supported in part by National Institutes of Health (NIH) (R01 EY023522, R01 EY029673, R01 EY030101, R01 EY030842, P30 EY001792); Richard and Loan Hill endowment; Unrestricted grant from Research to prevent blindness.

**Data Availability Statement:** The deep learning network ADC-Net is publicly available at https://github.com/dleninja/adcnet. Other data may be obtained from the authors upon reasonable request

**Figures:**

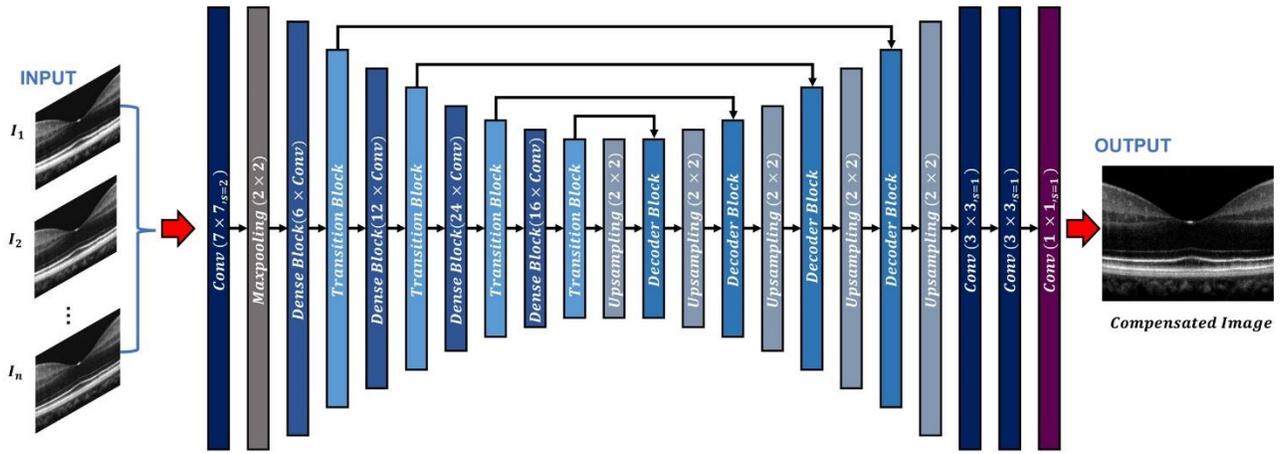

**Figure 1:** Overview of the ADC-Net architecture. Conv stands for convolution operation. The input section is comprised of OCT B-scans compensated by different second-order coefficients (n = 3, 5, 7, and 9 for three, five, seven, and nine input channels models respectively. A single image is used as input for the single input channel model). The output image is a corresponding fully compensated OCT B-scan.

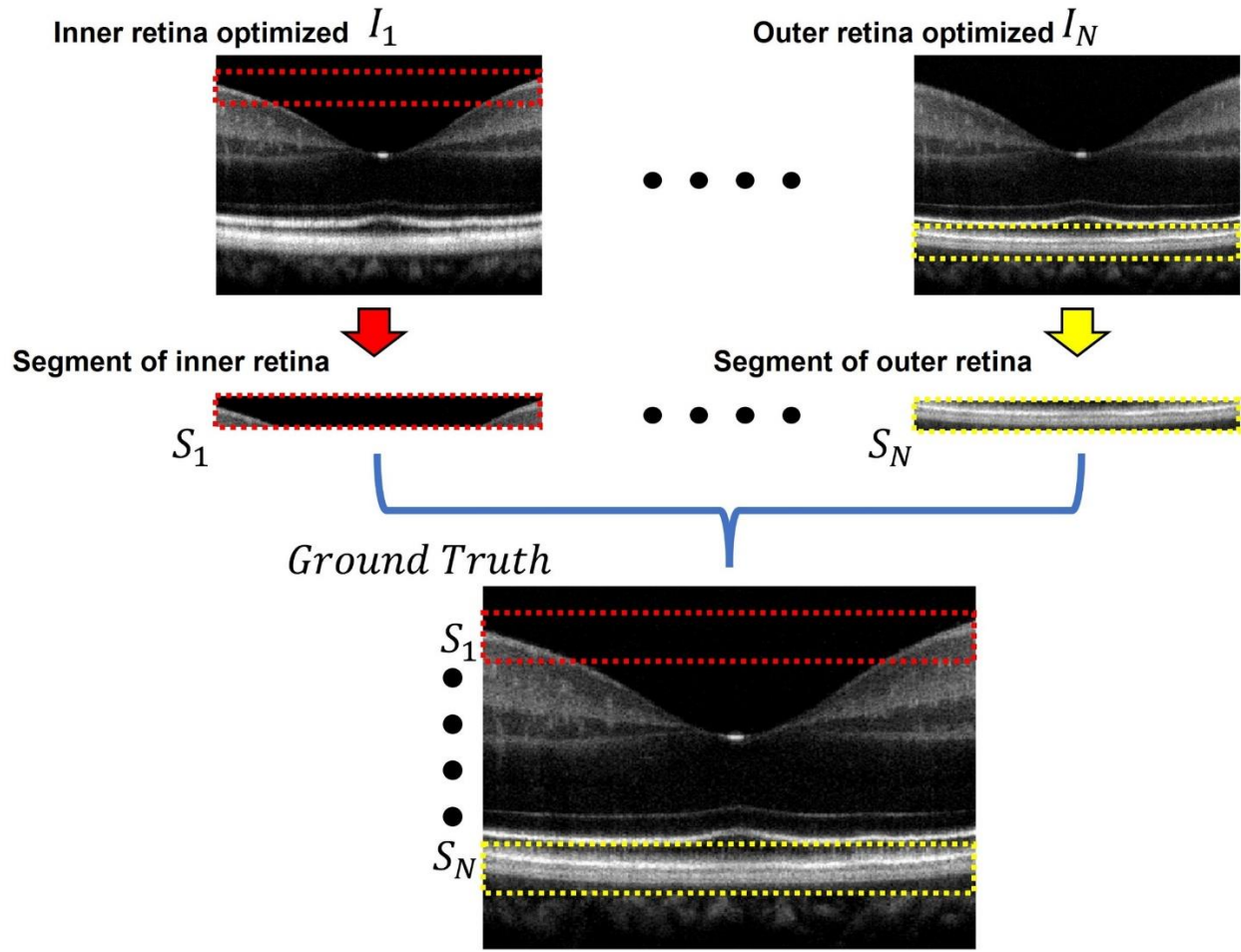

**Figure 2:** Ground truth preparation. An array of images ranging from $I_1$ to $I_N$ were utilized to prepare the ground truth image. Each image was compensated by a single second-order coefficient ranging from $C_1$ to $C_N$. Well compensated layers from each image were extracted and stitched together to form the ground truth image. In $I_1$ the inner retinal segment (demarcated by the dashed red box) is better compensated compared to the inner retinal segment in $I_N$. On the other hand, the outer retinal segment in $I_N$ (demarcated by the dashed yellow box) is better compensated compared to that of $I_1$. The ground truth was prepared by stitching the inner retinal segment $S_1$ from $I_1$ and outer retinal segment $S_N$ from $I_N$. The adjacent layers were similarly extracted and stitched from the subsequent images between $I_1$ and $I_N$.

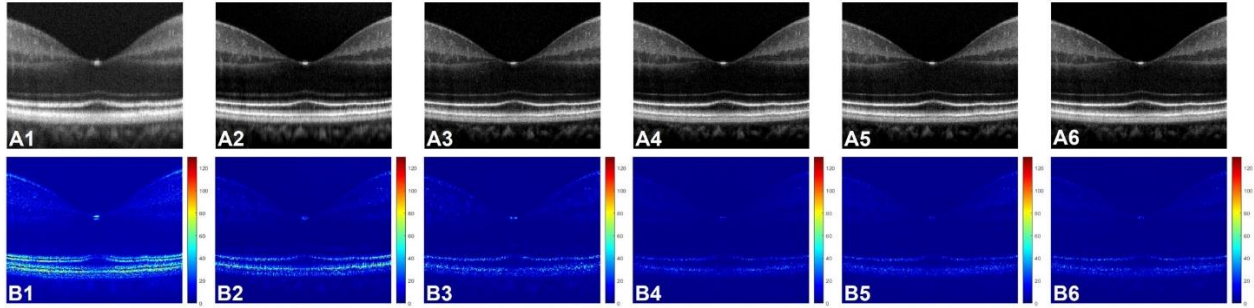

**Figure 3:** (A) Representative OCT of a human retina with no dispersion compensation (A1), and dispersion compensated by single (A2), three (A3), five (A4), seven (A5), and nine (A6) input channels models. (B) Corresponding differential intensity images which were generated by computing the pixel-to-pixel intensity differences between the ground truth image and the images in A1-A6, respectively.

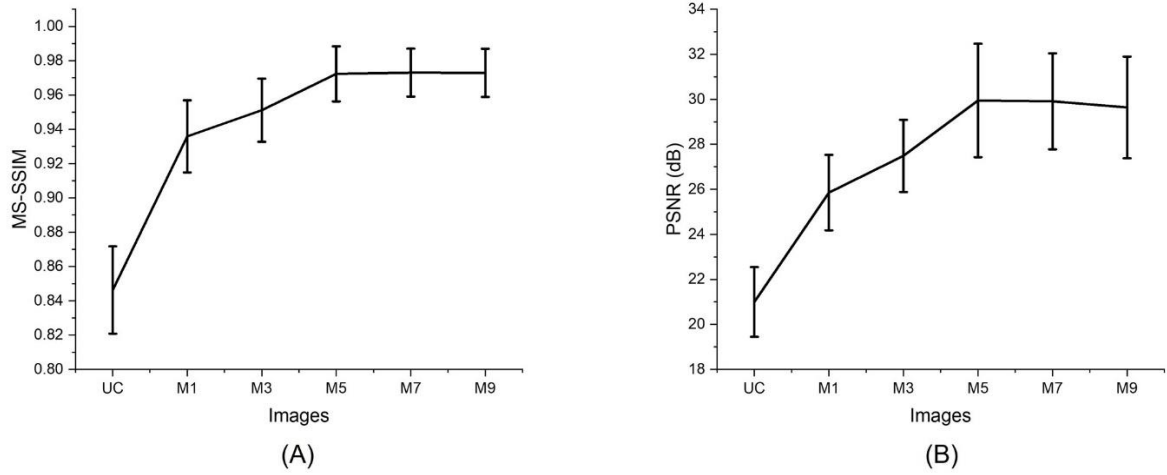

**Figure 4:** Quantitative evaluation of MS-SSIM (A) and PSNR (B). Average values are used for the graphical representation. UC represents the uncompensated images; M1, M3, M5, M7, and M9 represent the images acquired from ADC-Net with the single, three, five, seven, and nine input channels models.

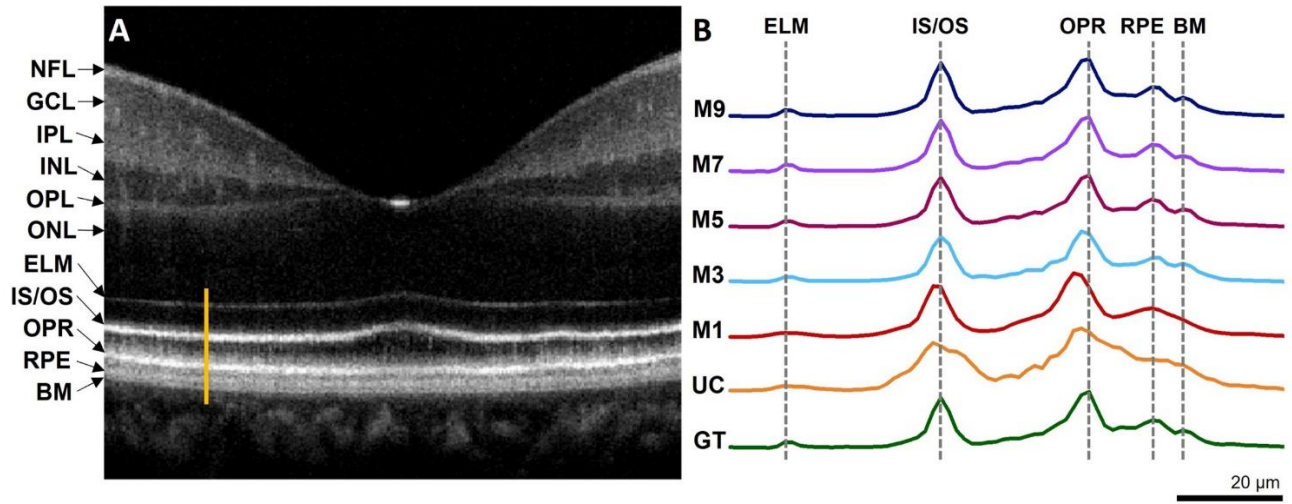

**Figure 5:** Intensity profile analysis of the outer retinal bands. (A) OCT B-scan of a human retina at the macula region with different retinal layers labeled accordingly. The bright orange line represents the A-line segment of the outer retina at the parafovea region where the intensity profiles in Figure 5(B) were computed. (B) Intensity profiles generated from different image models. GT: ground truth; UC: uncompensated; M1: single input channel; M3: three input channels; M5: five input channels; M7: seven input channels; M9: nine input channels. NFL: nerve fiber layer; GCL: ganglion cell layer; IPL: inner plexiform layer; INL: inner nuclear layer; OPL: outer plexiform layer; ONL: outer nuclear layer; ELM: external limiting membrane; IS/OS: inner segment/outer segment junction; OPR: outer segment PR/PRE complex; RPE: retinal pigment epithelium. BM: bruch's membrane.